\documentclass[preprint,showpacs,aps,amssymb,floatfix,prd,amsmath,preprintnumbers]{revtex4}
\usepackage{epstopdf}
\usepackage{capt-of}
\usepackage{float}
\usepackage{graphicx}  
\usepackage{dcolumn}   
\usepackage{bm}

\begin{document}
\title{Wormhole model with a hybrid shape function in $f(R,T)$ gravity}

\author{Parbati Sahoo$^{1}$\footnote{Email: sahooparbati1990@gmail.com }, Sanjay Mandal$^{2}$\footnote{Email: sanjaymandal960@gmail.com},
P.K. Sahoo$^{2}$\footnote{Email: pksahoo@hyderabad.bits-pilani.ac.in}}\
  
\affiliation{$^{1}$ Department of Mathematics, National Institute of Technology, Calicut, Kerala-673601, India}
\affiliation{$^{2}$Department of Mathematics, Birla Institute of
Technology and Science-Pilani, Hyderabad Campus, Hyderabad-500078,
India}

\begin{abstract}
\textbf{Abstract:} In the present article we propose a new hybrid shape function for wormhole (WH)s in the modified $f(R,T)$ gravity. The proposed shape function satisfied the conditions of WH geometry. Geometrical behavior of WH solutions are discussed in both anisotropic and isotropic cases respectively. Also, the stability of this model is obtained by determining the equilibrium condition. The radial null energy condition and weak energy condition are validated in the proposed shape function indicating the absence of exotic matter in modified $f(R,T)$ gravity.
\end{abstract}



\maketitle


\section{Introduction}

The Einstein field equations allow simple and interesting solutions for topological passage through hypothetical tunnels connecting two asymptotically flat portion of the same universe or two asymptotically flat universe. These solutions have been paid a lot of attention to their physical effects and gravitational properties. The solution would act as shortcut passage between two distant region of space-time. They can be used for constructing time machine, for which a stable traversable WH is required \cite{Morris/1988,Throne/1988,Frolov/1990,Visser/1993,Bertolami/2012}.

In the early stage of General Relativity (GR), the WH-like solution was obtained in  \cite{Flamm/2015}. Einstein and Rosen obtained an exact solution of the field equations in their seminal work \cite{Einstein/1935}. These solutions corresponding to a space-time of two spread sheet joined by a tunnel or bridge and related to Flamm solution \cite{Gibbons/2017}. Later the word `wormhole' was used by Wheeler and Misner \cite{Wheeler/1957}. Further theses WHs contains certain kind of matter, called `exotic' matter, which act an important role to its geometry. Moreover, the existence of traversable WH is  only possible in the presence of that exotic matter which may have negative energy and would violates all energy conditions \cite{Eiroa/2010,Anchordoqui/1997}. There is another way to relocate this violation of energy conditions, for example; the gravitational effects of the dark energy or phantom energy \cite{Thibaut/2017,Sushkov/2005,Lobo/2005} which explains the accelerated expansion of the universe, are also violates energy conditions and shows the existence of such kind of matter field seems to be suitable. Several works related to phantom energy and its roles in WH geometry are investigated widely in literature \cite{Wang/2016,Sahoo/2019a,Sahoo/2019b} and see the references therein. 

In this work we have considered a static spherically symmetric WH metric with Schwarzschild coordinates $(t,r,\theta, \phi)$ is \cite{Morris/1988, visser/1995}
\begin{equation}\label{1}
ds^2=-e^{2\Phi(r)}dt^2+\frac{dr^2}{1-\frac{b(r)}{r}}+r^2d\Omega^2,
\end{equation}
where $d\Omega^2=d\theta^2+\sin^2\theta d\phi^2$, $\phi(r)$ and $b(r)$ are called the shape function and the redshift function respectively. $b(r)$ determines the spatial shape of the WH. The radial coordinate $r$ lies between $r_0\leq r <\infty$, where $r_0$ is known as the throat radius. WHs are purely theoretical due to lack of observation. So, we don't have any fixed formula/function for it's geometry and equation of states. But we can't consider the shape function or redshift function arbitrarily as they must obey some metric conditions which are given below.

The shape function $b(r)$ in metric (\ref{1}) should obey the throat condition, at the throat $r=r_0$,  $b(r_0)=r_0$ and at $r>r_0$ i.e. out of the throat, $1-\frac{b(r)}{r}>0$. For the flaring out condition at the throat, $b(r)$ has to obey $b'(r_0)<1$. In order to maintain an asymptotically flatness of the spacetime
geometry, it require the limit $\frac{b(r)}{r}\rightarrow 0 \ \ \ \text{as}\ \ \ \ \vert r\vert\rightarrow \infty$. To avoid an event horizon issues, the redshift function has to finite everywhere and in this model, we have chosen a constant redshift function for obtaining a WH solution. In addition, we have introduced a shape function and it follows all the metric conditions as discussed in the next section.

In the process of constructing a physically reliable model, the problem is to find the suitable contenders for exotic matter which has never been done. Due to the problematic nature of exotic matter (it's distribution), it is useful to minimize the usage of exotic matter. Therefore, \cite{Moraes/2017} are added some extra degrees of freedom in it's fundamental level to minimize it in modified gravity theories. Thereafter, many WH models are widely investigated in the framework of modified gravity theories.\\
For instance, the existence of WH solution with non-exotic matter have already been obtain in multimetric repulsive gravity model \cite{Hohmann14}, in higher order curvature gravity \cite{sahoo188,harko13}, in a trace of energy-momentum tensor squared gravity \cite{moreas18}, respectively. In $f(R,T)$ gravity \cite{Zubair16}, where the authors have investigated the existence of WH solution in presence of different matter content (anisotropic, isotropic, and barotropic) and found the existence of the WH solution without exotic matter in few regions of space time.  Accordingly, a static spherically symmetric WH solution is investigated with diffrent matter contents and specific choices of redshift function in the framework of $f(T,T_G)$ gravity \cite{Sharif18}, and $f(G,T)$ gravity \cite{Sharif17}, respectively. On the other hand, the modified $f(R,T)$ gravity is widely used in investigation of emergence of compact stellar objects in \cite{Yousaf18a}, and the causes of irregular energy density and irregularity factors for a self-gravitating spherical star evolving with an imperfect fluid in \cite{Yousaf16b}. In addition, the construction of charged cylindrical gavastar-like structure is analyzed in the framework of $f(R,T)$ gravity \cite{Yousaf20c}, and some WH solutions are explored with extra matter contents in it \cite{Yousaf17d} ( also see \cite{Yousaf17e, Yousaf18f} and references therein). However, all these literatures and the references therein are motivate us and provide ideas to focus on more new findings.  

In this paper, we study static spherically symmetric WH metric in $f(R,T)$ gravity with constant redshift function. The paper is organized as follows:  we discuss the necessary metric conditions of shape function $b(r)$ and redshift function $\Phi (r)$ in first section. In section II, we derive the field equations and it's solutions in $f(R,T)$ gravity and analyze the behaviors of shape function and energy conditions. In section III, we discuss the equilibrium condition of the WH solutions. Lastly, we have discussed the results in section V.

\section{Field equations and its solution in $f(R,T)$ gravity}

The total action in the $f(R,T)$ theory of gravity reads \cite{harko/2011}
\begin{equation}\label{e1}
S=\frac{1}{16\pi}\int d^{4}x\sqrt{-g}f(R,T)+\int d^{4}x\sqrt{-g}\mathcal{L}_m,
\end{equation}
where $f(R,T)$ is an arbitrary function of Ricci scalar, $R=R^i_j$ and $T=T^i_J$ is the trace of the stress-energy tensor of the matter with $g$ being the metric determinant. $\mathcal{L}_m$ is the matter Lagrangian density related to stress-energy tensor as
\begin{equation}\label{e2}
T_{ij}= - \frac{2}{\sqrt{-g}} \frac{\delta(\sqrt{-g}\mathcal{L}_{m})}{\delta
g^{ij}}.
\end{equation}
Since, the matter Lagrangian density $\mathcal{L}_m$ depends only on the metric $g_{ij}$, we have
\begin{equation}\label{e3}
T_{ij}= g_{ij}\mathcal{L}_{m} - 2\frac{\partial \mathcal{L}_{m}}{\partial g^{ij}}.
\end{equation}
By varying the action $S$ given in (1) with respect to metric $g_{ij}$ provides the $f(R,T)$ field equations 
\begin{multline}\label{e4}
f_R(R,T)\left(R_{ij}-\frac{1}{3} Rg_{ij}\right) + \frac{1}{6}f(R,T)g_{ij} \\=8\pi G \left(T_{ij}-\frac{1}{3}Tg_{ij}\right)-f_T(R,T)\left(T_{ij} -\frac{1}{3}Tg_{ij}\right)\\-f_T(R,T)\left(\theta_{ij}-\frac{1}{3}\theta g_{ij}\right)+\nabla_i\nabla_jf_R(R,T).
\end{multline}
Here, the notations are $f_R(R,T)=\delta f(R,T)/\delta R$ and $f_T(R,T)=\delta f(R,T)/\delta T$ respectively and
\begin{equation}\label{e5}
\theta_{ij}=g^{ij}\frac{\delta T_{ij}}{\delta g^{ij}}.
\end{equation}
We have considered a natural form the matter Lagrangian for this model is $\mathcal{L}_m=-\rho$, where $\rho$ is the energy density. In the choice of matter Lagrangian, one can consider with any form of matter components like pressure (p) or energy density ($\rho$), which
can be found in \cite{harko/2011}. Also the natural choice of $\mathcal{L}_m= -\rho$ is more generic and does not imply the vanishing of the extra force. On the other hand, by considering the form $\mathcal{L}_m= -\mathcal{P}$, where $\mathcal{P}$ is the total pressure, the extra force vanishes \cite{Bert/2008}. This is due to the extra force depends on the form of matter Lagrangian $\mathcal{L}_m$ \cite{Harko2014}. Here, with the given choice of $\mathcal{L}_m$, equation (\ref{e5}) can be written as
\begin{equation}\label{e6}
\theta_{ij}=-2T_{ij}-\rho g_{ij}.
\end{equation}
Let us assume the linear form of $f(R,T)$ gravity, i.e. $f(R,T)=R+2f(T)$, where $f(T)$ is an arbitrary function of the trace of energy momentum tensor $T$ only, employed as $f(T)=\lambda T$, where $\lambda$ is constant. It is worth to point out that the $f(R, T)$ gravity theories in this form, can be reduced to GR with the choice of $\lambda=0$. That means only the material sector of GR is modified here. However, the extended theories of gravity allows one to modify the material sector without disturbing the geometrical sector of GR. This linear form may provide a very close comparison between GR and modified gravity theories.\\

The $f(R,T)$ gravity field equations (\ref{e4}) with (\ref{e6}) takes the form
\begin{equation}\label{7}
R_{ij}-\frac{1}{2}Rg_{ij}=8\pi G T_{ij}+2f'(T)T_{ij}+[2\rho f'(T)+f(T)]g_{ij}.
\end{equation}
Assuming $8\pi G \equiv 1$ and $f(T)=\lambda T$, the above equation can be written as
\begin{equation}\label{8}
R_{ij}-\frac{1}{2}Rg_{ij}=(1+2\lambda) T_{ij}+(2\rho+T)\lambda g_{ij}.
\end{equation}

The energy-momentum tensor for an anisotropic fluid satisfying the matter content is given as
\begin{equation}\label{14}
T^i_{j}=(\rho +p_t)u^iu_j-p_tg^i_j+(p_r-p_t)x^ix_j
\end{equation}
Here, $\rho(r)$ is the energy density, $p_r(r)$ and $p_t(r)$ are the radial and lateral pressure measured orthogonally to the radial direction respectively. $u_i$ and  $x_i$ are the four-velocity vector and radial unit four vector defined with the relation $u^iu_i=1$ and $x^ix_i=-1$ respectively. 


 The trace $T$ of the energy momentum tensor (\ref{14}) turns out to be $T=-\rho+p_r+2p_t$. The field equations (\ref{8}) for the metric (\ref{1}) with (\ref{14}) are as follows
\begin{equation}\label{15}
\frac{b'}{r^2}=(1+\lambda)\rho-\lambda(p_r+2p_t),
\end{equation}
\begin{equation}\label{16}
-\frac{b}{r^3}= \lambda \rho+(1+3\lambda)p_r+2\lambda p_t,
\end{equation}
\begin{equation}\label{17}
\frac{b-b'r}{2r^3}= \lambda \rho+\lambda p_r+(1+4\lambda) p_t.
\end{equation}
The above set of field equations admits the general solution
\begin{equation}\label{18}
\rho=\frac{b'}{r^2(1+2\lambda)},
\end{equation}
\begin{equation}\label{19}
p_r= - \frac{b}{r^3(1+2\lambda)},
\end{equation}
\begin{equation}\label{20}
p_t= \frac{b-b'r}{2r^3(1+2\lambda)}.
\end{equation}

In this work, we defined new shape function $b(r)$ as follows:
\begin{equation}\label{21}
b(r)=r_0^m e^{r_0-r}r^{1-m}.  
\end{equation}
Motivated by \cite{Lobo/2009} we have considered $m=2$, and plotted the corresponding graphs as given below.  

\begin{figure}[H]
\centering
\includegraphics[scale=0.5]{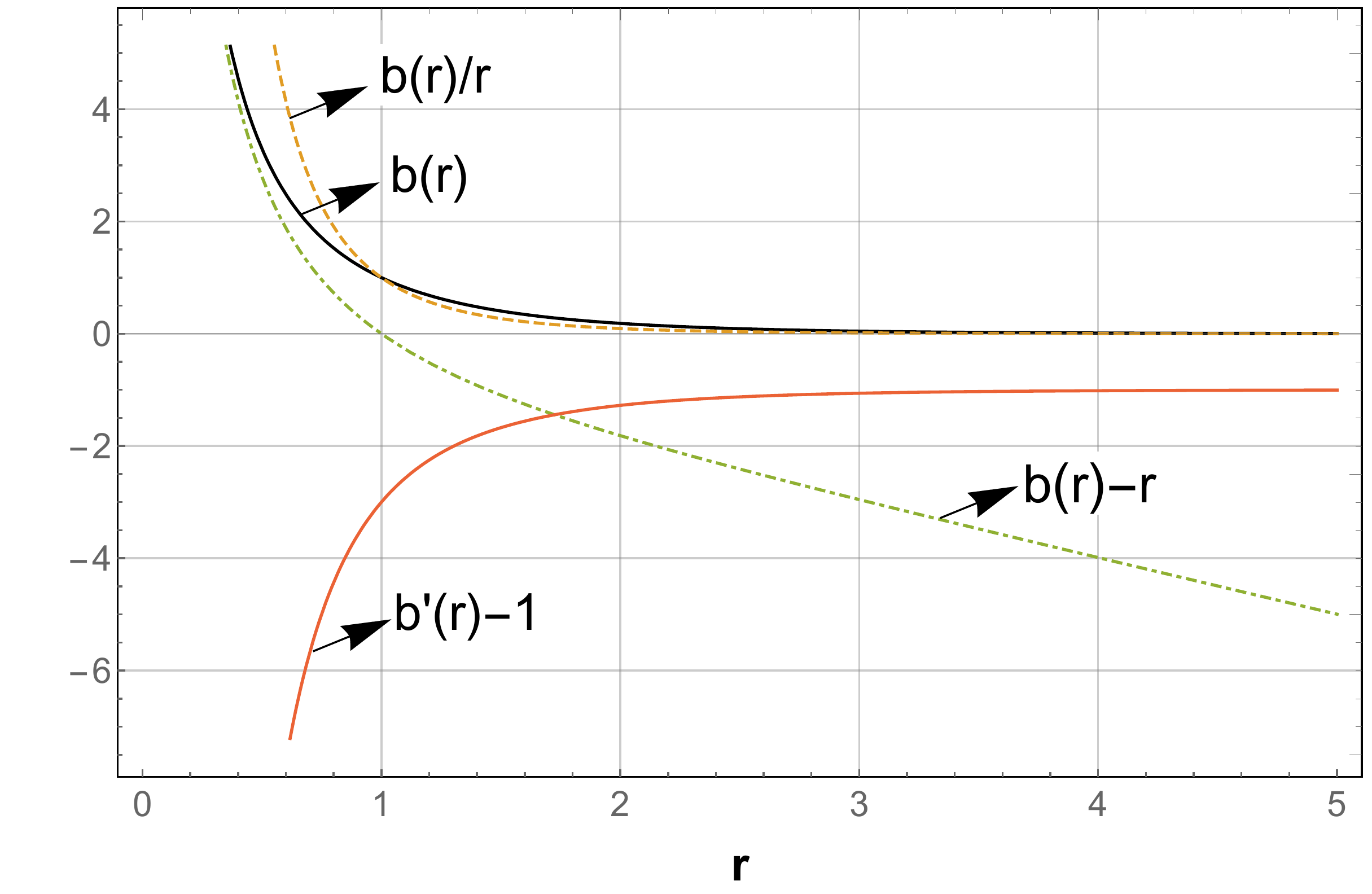}
\caption{Behavior of shape function with $r_0=1$.}
\label{fig1}
\end{figure}

\begin{figure}[H]
\centering
\includegraphics[scale=0.5]{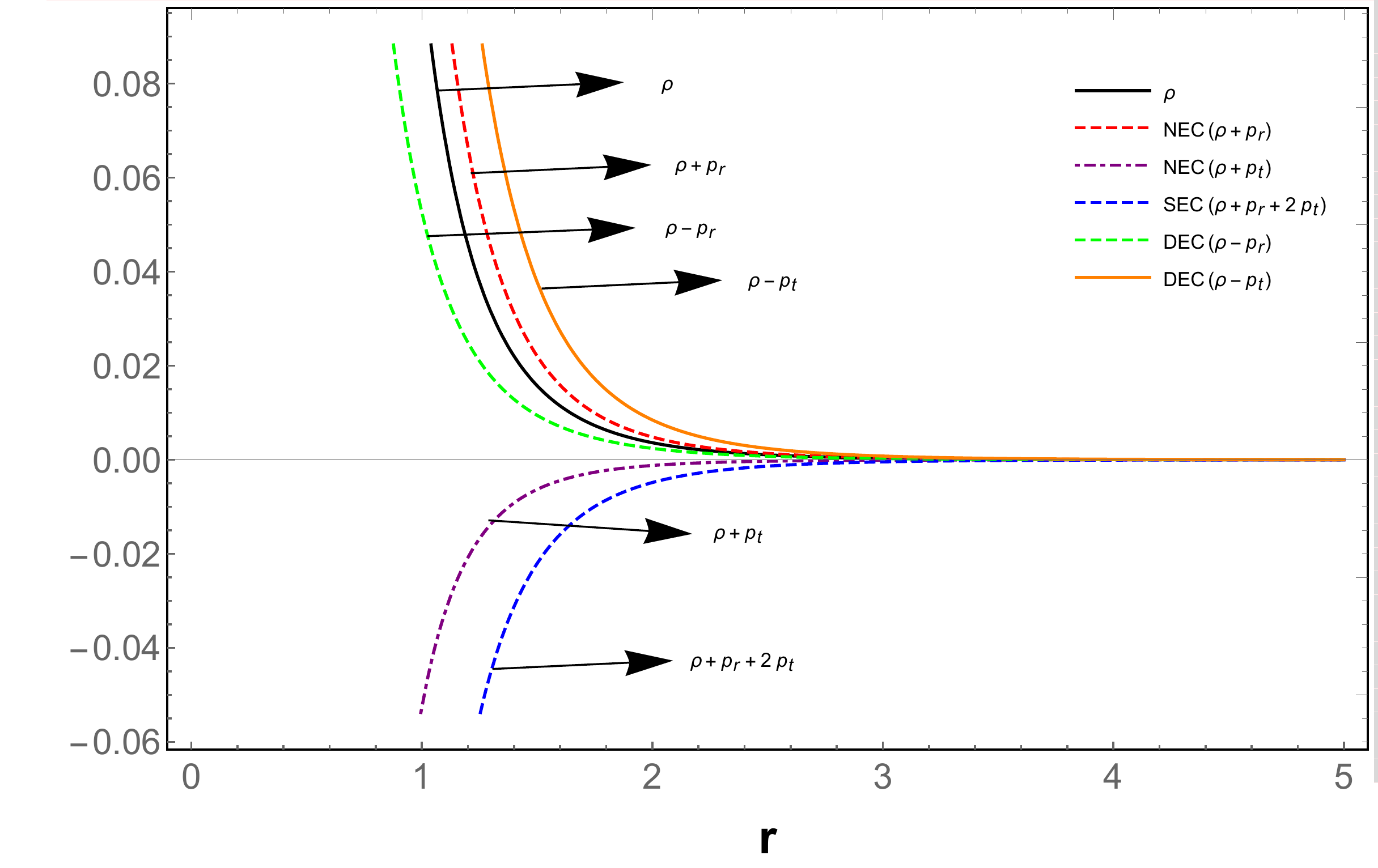}
\caption{Behavior of $\rho$ and ECs with $r_0=1$ and $\lambda=-10$.}
\label{fig2}
\end{figure}
 
The metric conditions discussed in first section are obeyed by the proposed shape function (see Fig. \ref{fig1}). Also, the solutions obtained from field equations by using eqn. (\ref{21}) are agreeing with physical properties of the WH. One can observe that, Fig. \ref{fig2} represents the behavior of all energy conditions along with energy density for specified range of $\lambda \in [-10,-1]$. It can be seen that the null energy condition violates in tangential coordinate, while validates in radial coordinate. Moreover the strong energy condition  violates and provides an extra evidences for accelerated expansion of universe. In case of dominant energy condition, it satisfies in both radial and tangential coordinates. It gives a chance to go with the stability of this model. However, the EoS parameter, $\omega_r=\frac{p_r}{\rho}$ in this case is independent of $\lambda$. Therefore, any range of $\lambda$ does not affects the behavior of EoS parameter. 

\section{Equilibrium condition}

Here, we have focused on the equilibrium configuration of the
WH solutions and it addressed by using the generalized
Tolman-Oppenheimer-Volkoff (TOV) equation given by 
\begin{equation}\label{22}
\frac{dp_r}{dr}+\frac{a'(\rho+p_r)}{2}+\frac{2}{r}(p_r-p_t)=0,
\end{equation}
for the metric tensor $g_{ij}=\left\{-e^{a(r), \left(1-\frac{b(r)}{r}\right)^{-1}, r^2, r^2 \sin \theta^2}\right\}$ given in equation (\ref{1}), where $a(r)=2\Phi (r)$.

This equation provides the equilibrium picture of static WH solutions through three forces namely, gravitational force $F_{gf}$, anisotropic force $F_{af}$ and hydrostatic force $F_{hf}$ . The gravitational force appears as the result of
gravitating mass, anisotropic force arises due to anisotropy of the system and hydrostatic force occurs as a result of hydrostatic fluid.
Accordingly, the gravitational, hydrostatic, and anisotropic force due to anisotropic matter distribution are defied as follows:
\begin{eqnarray}\label{23}
F_{gf}=-\frac{a'(\rho+p_r)}{2},\\
F_{af}=\frac{2(p_t-p_r)}{r},\\
F_{hf}=-\frac{dp_r}{dr}.
\end{eqnarray}
It is required that $F_{gf}+F_{af}+F_{hf}=0$  must hold for the WH solutions to be in equilibrium.
Since we considered a constant redshift function i.e. $\Phi(r)=\text{constant}$, we have $F_{g f} = 0$ and hence the equilibrium condition reduces to the following form

$F_{af}+F_{hf}=0$

Here, we calculate the values of $F_{hf}$ and $F_af$ as 
\begin{eqnarray}\label{24}
F_{hf}=-\frac{e^{1-r} (r+4)}{(2 \lambda +1) r^5},\\
F_{af}=\frac{2 e^{1-r} (r+3)}{(2 \lambda +1) r^5}.
\end{eqnarray}

 \begin{figure}[H]
\centering
\includegraphics[scale=0.5]{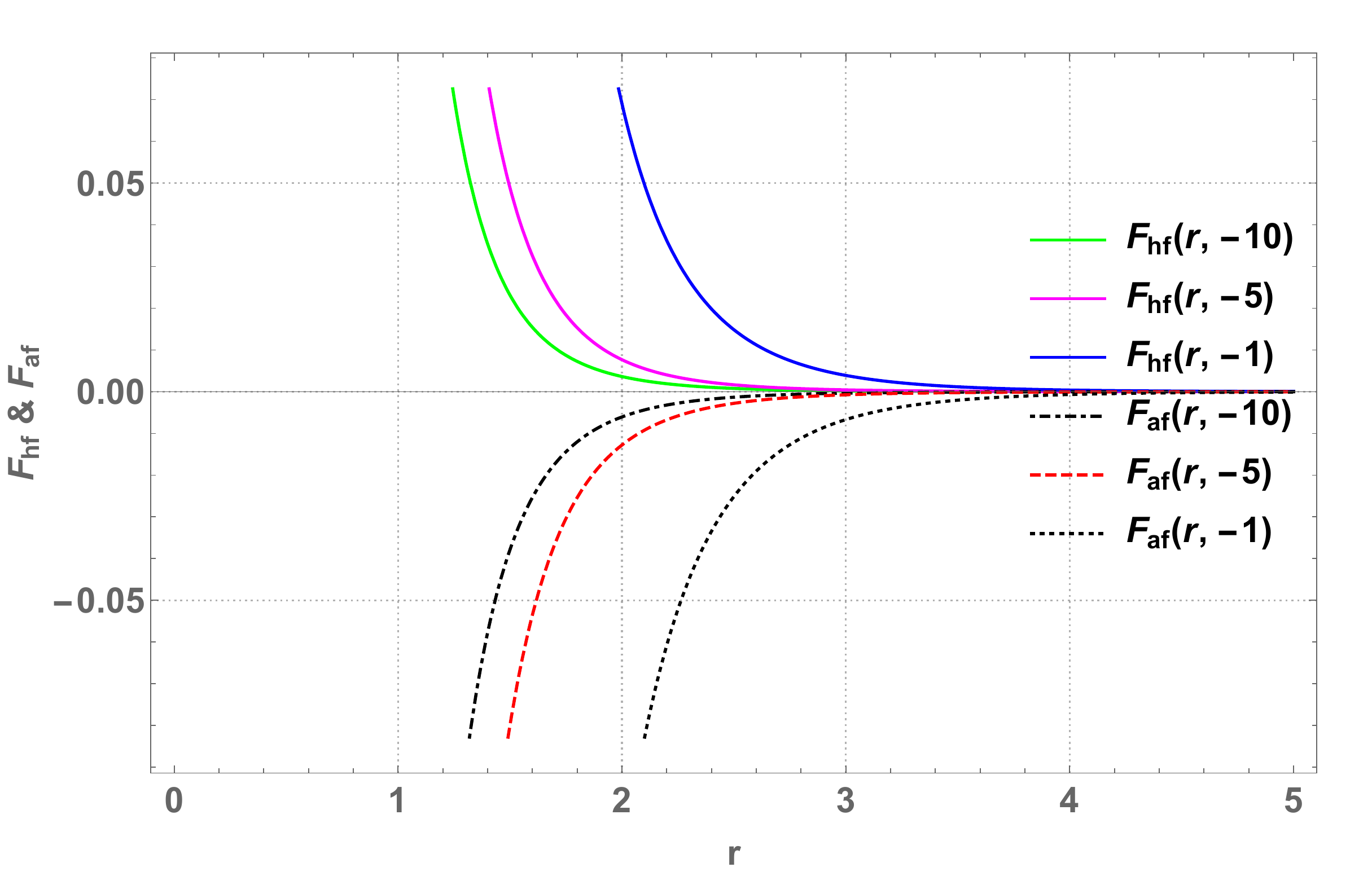}
\caption{Behavior of $F_{hf}$ and $F_{af}$ with $r_0=1$ and different $\lambda$.}
\label{fig9}
\end{figure}
The graphical illustration of $F_{hf}$ and $F_{af}$ forces versus $r$ for three representative values of $\lambda$ are depicted in Fig. \ref{fig9}. It defines the attractive geometry. This behavior can be studied through the dimensionless anisotropy parameter \cite{Cattoen/2005,Lobo/20133} for anisotropic pressures. Since $\rho>0$, the relation $\frac{p_t-p_r}{\rho}$ represents a force due to the anisotropic nature of the WH model. Geometry is attractive if $p_t<p_r$, i.e. $\Delta <0$, and repulsive if $p_t>p_r$, i.e. $\Delta>0$. The fluid is isotropic for  $\Delta=0$, i.e. $p_t=p_r$.
In the present case, we have obtained 
\begin{equation}\label{25}
\Delta= \frac{p_t-p_r}{\rho}=-\frac{3+r}{1+r},
\end{equation}
where the value of $\Delta <0$ for all positive values of $r$. It confirms the attractive nature of geometry of this present model for anisotropic fluid matter.

\section{Discussion}

In the present work, we have studied an exact WH solution with a newly proposed shape function. As WHs are classified into three categories in general, such as ordinary WH, traversable WH, and thin-shell WH. Ordinary WHs are based on only satisfaction of NEC violations and usually not asymptotically flat, singular, and consequently non-traversable respectively. In case of traversable WH, we can obtain this by an appropriate choice of redshift function or shape function. In addition, one can analyze the traversability conditions of the WH and its stability. Later, the thin shell WH can be constructed theoretically by geodesically complete traversable WH with a shell placed in the junction surface by using the so-called cut-and-paste technique. In this model, we focused on a special kind of shape function to construct WH solution which provides an attractive geometry. We have started the WH construction by considering an anisotropic fluid and with an appropriate choice of shape function. We found a stable WH solution with attractive geometry. Also, the null energy condition violated in the tangential coordinate while obeys in the case of radial pressure as depicted in the fig \ref{fig2}. In section III, we have discussed the equilibrium conditions by using the TOV equation and found the positive hydrostatic force and negative anisotropic force. Both are satisfying the TOV equation and provides an idea of attractive geometry. This is validated through the behavior of anisotropic parameter given in eqn. (\ref{25}). 
On the the hand, for an isotropic distribution of pressures (i.e. $p_r=p_t=p$), the WH solution obtained through the defined shape function has some interesting behaviors. First of all, from eqn. (\ref{15}-\ref{17}), we have obtained the values are as follows
\begin{eqnarray}
\rho=-\frac{e^{1-r} (4 \lambda +(7 \lambda +2) r+2)}{2 \left(8 \lambda ^2+6 \lambda +1\right) r^4},\\
p=\frac{e^{1-r} (4 \lambda +3 \lambda  r+r+2)}{2 \left(8 \lambda ^2+6 \lambda +1\right) r^4}.
\end{eqnarray}    
Furthermore, to discuss about the equilibrium condition, the gravitational and anisotropic force are zero respectively, while the hydrostatic force is obtained as
\begin{equation}
F_{hf}=\frac{e^{1-r} \left(16 \lambda +(3 \lambda +1) r^2+(13 \lambda +5) r+8\right)}{2 \left(8 \lambda ^2+6 \lambda +1\right) r^5}.
\end{equation}
Here, the hydrostatic force is found as negative for the given value of $\lambda$ and diverges to zero as to follow the TOV equation. This nature of hydrostatic force indicates a repulsive geometry. Furthermore, we have pointed in the last paragraph of section-II that the EoS parameter in anisotropic case has no role for any value of $\lambda$. But in isotropic case, the EoS parameter, $\omega_{r(isotropic)}=\frac{p}{\rho} >-1$ for the given range of $\lambda \in [-10, -1]$. It shows the quintessence phase, while the phantom phase, $\omega_{r(isotropic)} <-1$ occurs at the some certain range of $\lambda$, i.e. $-0.3<\lambda<-0.26$. This range of $\lambda$ does support the positivity of energy density in anisotropic case. To maintain the uniformity of the model solutions, we restricted the range of $\lambda \in [-10, -1]$. As it known that the phantom phase supports the existence of exotic matter at the throat of WH, which is responsible for NEC violation. The derived model in this case avoids the existence of exotic matter at the throat of WH by absorbing the quintessence phase. In this context, one can refer these references \cite{Harko2013,Zubair2019,Samanta2019} (and references therein) to study more details about the various modified gravity WH models without exotic matter. \\ 

\begin{figure}[H]
\centering
\includegraphics[scale=0.5]{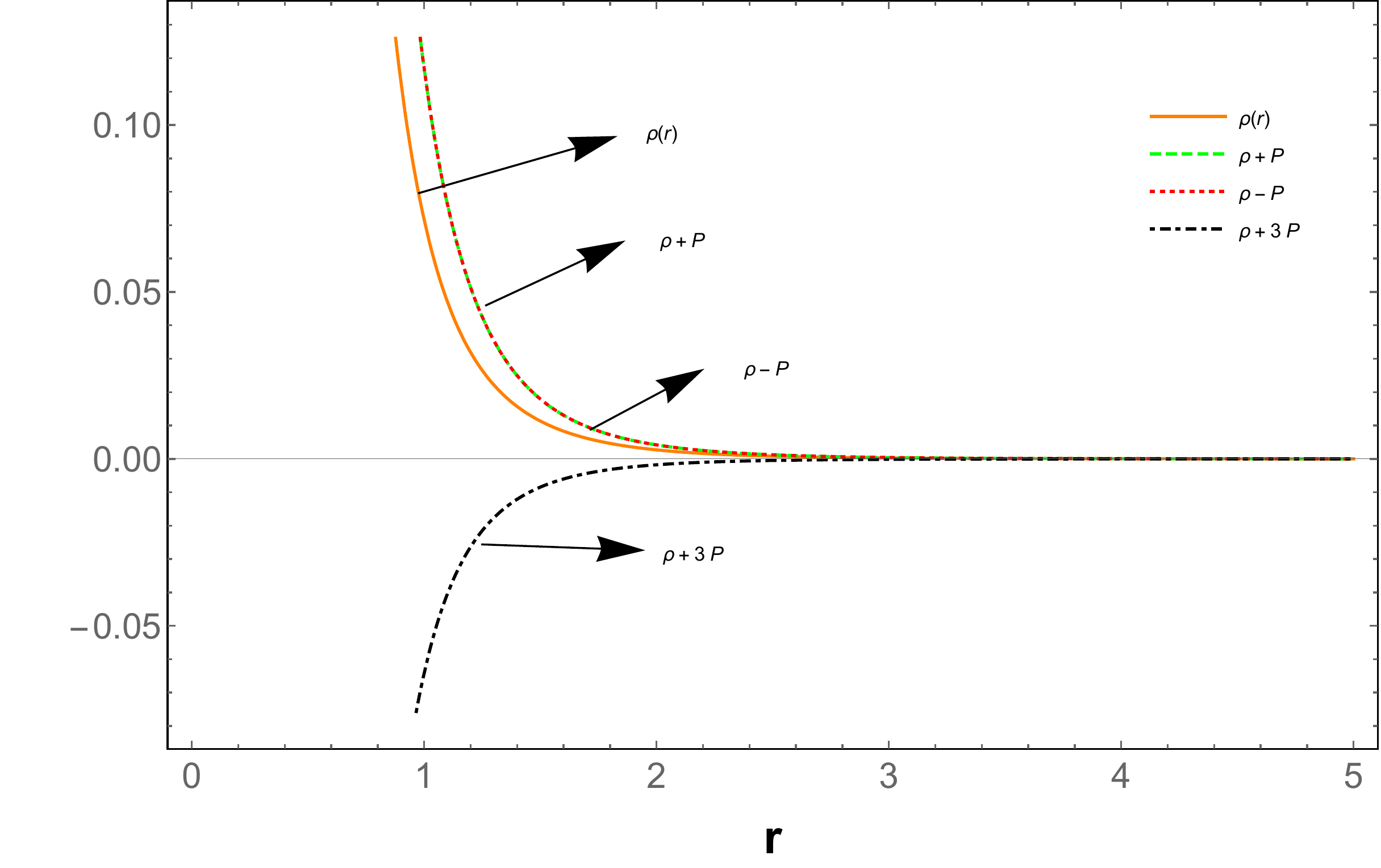}
\caption{Behavior of energy conditions with $r_0=1$ and $\lambda=-10$.}
\label{fig4}
\end{figure}
Furthermore, the energy conditions are obtained in the isotropic case are represented in Fig. \ref{fig4}.
It can be observed that, the NEC for isotropic distribution of pressures  is validating. We can say that the use of the shape function in eqn. (\ref{21}) may reduce the presence of exotic matter in the WH throat for this isotropic case. Moreover, we found the SEC violates for both anisotropic and isotropic case, which indicates the accelerated expansion of the universe due to modified $f(R,T)$ gravity. Since, the rate of expansion is characterized by the isotropic and homogeneous matter distribution, it is worth pointing out that, in late time accelerating universe, the presence of exotic matter can be reduced and an exact WH solution without exotic matter can be obtained in the framework of modified gravity theories.\\
At the end, we can conclude by saying that the existence of a physically acceptable WH is possible by using the newly proposed shape function in the framework of modified gravity theories. Moreover in future work, we are focusing on how far this shape function supports the observational consequence of the WH geometry.   

\acknowledgments S.M. acknowledges Department of Science \& Technology (DST), Govt. of India, New Delhi,
for awarding Junior Research Fellowship (File No.
DST/INSPIRE Fellowship/2018/IF180676). PKS acknowledges DST, New Delhi, India for providing facilities through DST-FIST lab, Department of Mathematics, BITS-Pilani, Hyderabad Campus where a part of this work was done. We are very much grateful to the honourable referee and the editor for the illuminating suggestions that have significantly improved our work in terms of research quality and presentation.

\end{document}